\documentclass[a4paper,11pt]{article}

\usepackage{contribution}

\newcommand{\weblink}[2][]{%
    \ifthenelse{\equal{#1}{}}%
    {\textnormal{\url{#2}}}%
    {\textnormal{\href{#2}{#1}}}%
}

\def\beq{\begin{equation}}
\def\eeq#1{\label{#1}\end{equation}}
\def\eeqn{\end{equation}}

\def\beqa{\begin{eqnarray}}
\def\eeqa#1{\label{#1}\end{eqnarray}}
\def\eeqan{\end{eqnarray}}

\let\bar=\overbar

\def\Dslash{\not{\hbox{\kern-4pt $D$}}}
\def\dslash{\not{\hbox{\kern-2pt $\del$}}}

\def\msb{{\bar{\ssstyle M \kern -1pt S}}}

\newcommand{\contribution}[7][]{%
  \clearpage
  \thispagestyle{plain}
  \ifthenelse{\equal{#1}{}}
  {\hypersetup{pdftitle={#2}}}
  {\hypersetup{pdftitle={#1}}}
  \hypersetup{pdfauthor={{#3} {#4}}}
  {\centering\normalfont\LARGE\bfseries\sffamily #2 \par\nobreak}
  \lhead{}
  \chead{%
    \textit{\footnotesize XIV International Conference on Hadron Spectroscopy
      (\weblink[\textit{hadron2011}]{http://www.hadron2011.de}), 13-17 June 2011, Munich, Germany}%
  }
  \rhead{}
  \bigskip
  \begin{center}
    {#3} {#4}\ifthenelse{\equal{#6}{}}{}{\footnote{\weblink[#6]{mailto:#6}}}
    \ifthenelse{\equal{#7}{}}{}{#7} \\
    \textit{#5}
  \end{center}
  \bigskip
}

\renewcommand{\abstract}[1]{%
  \begin{center}
    \begin{minipage}{0.85\textwidth}
      \begin{footnotesize}
        #1
      \end{footnotesize}
    \end{minipage}
  \end{center}
  \bigskip
}

\begin{document}

%
%
%
%
%
{  


%

\contribution[Hadron Physics at KLOE and KLOE-2]  
{Hadron Physics at KLOE and KLOE-2}  
{Camilla}{Di Donato}  
{I.N.F.N. Sezione di Napoli \\
   Complesso Universitario M.S.A., Via Cintia ed.6\\
  I-80126 Napoli, ITALY}  
{didonato@na.infn.it}  
{on behalf of KLOE and KLOE-2 Collaborations}  
%

\abstract{%
The KLOE Collaboration completed the full data taking on March 2006, acquiring $2.5 ~fb^{-1}$ at the peak of the $\phi$ and other $240 ~pb^{-1}$ off-peak. A new Collaboration is working on a new project, called KLOE-2, to refine and extend the KLOE physics program.\\ We present here some preliminary and published results from the KLOE Collaboration on the pseudoscalar $\eta$ meson and the study of $\gamma \gamma$ processes, that are among the main points of the KLOE-2 physics program.
}
%

\section{Introduction}
The KLOE experiment has collected $2.5 ~fb^{-1}$ at the peak of the $\phi$ resonance at the $e^+e^-$ collider DA$\Phi$NE in Frascati. KLOE has performed several precision measurements, here we present the preliminary analysis of the $\eta \to \pi^+ \pi^- \gamma$ decay channel to study box anomaly and the  recently published branching ratio measurement of $\eta \to e^+ e^- e^+ e^-$ decay channel, never observed before. Pseudoscalar production at the $\phi$-factory associated to internal conversion of the photon into a lepton pair allows the measurement of the form factor $F(q_1^2=M(\phi)^2,q_2^2>0)$ of pseudoscalar mesons in the kinematical region of interest for the VMD model: a preliminary study of $\phi \to \eta e^+ e^-$ is based on $739 ~pb^{-1}$, using the $\eta \to \pi^+ \pi^- \pi^0$ final state.\\ 
From a sample of $240 ~pb^{-1}$ taken off the $\phi$ resonance, a preliminary analysis of the $e^+ e^- \to e^+ e^- \eta$ process, without $e^{\pm}$ tagging in the final state has been performed. The same data set has been used to search for the $f_0(600)$ produced in $\gamma \gamma$ interactions via the reaction $e^+ e^- \to e^+ e^- \pi^0 \pi^0$.\\ 
The KLOE detector is being upgraded with small angle tagging devices, to detect both high and low energy $e^{\pm}$ in $e^+ e^- \to e^+ e^- X$ events. The inner tracker and small angle calorimeters are scheduled to be installed in a subsequent step, providing wider acceptance for both charged particles and photons. This is the new KLOE-2 project \cite{AmelinoCamelia:2010me}: the detector is successfully rolled in the new DA$\Phi$NE interaction region, with a new beam crossing scheme allowing for a reduced beam size and increased luminosity. The main goal of KLOE-2 is to collect an integrated luminosity of about 
$20~fb^{-1}$ in 2-3 years in order to refine and extend the KLOE physics programme.
\section{The Pseudoscalar $\eta$ meson}
The decays $\eta, \eta^{\prime} \to \pi^+\pi^-\gamma$ are supposed to get contribution from the anomaly accounted by the Wess Zumino Witten term into the Chiral Perturbation Theory Lagrangian \cite{Benayoun:2003we}. Those anomalous processes, known as box anomalies,
proceed via a vector meson resonant contribution (VDM) and maybe via a direct term. The presence of this direct term affects the partial width value in the case of the $\eta \to \pi^+\pi^-\gamma$ and the dipion invariant mass distribution, in the case of $\eta^{\prime} \to \pi^+\pi^-\gamma$. \\
A comparison of the experimental $M_{\pi^+ \pi^-}$ spectra and partial width for $\eta$, $\eta^{\prime}$ meson with theoretical predictions is mandatory to clarify the role of non-resonant contribution to the processes.
The $\eta \to \pi^+ \pi^- \gamma$ decay has been measured in 1970 by Gormley et al.~(7250 events)~\cite{Gormley:1970qz} and in the 1973 by Layter et al.~(18150 events)~\cite{Layter:1973ti.Thaler:1973th}. Theoretical papers trying to combine the two measurements found discrepancies in data treatment and problems in obtaining consistent results~\cite{Benayoun:2003we}. In 2007 CLEO Coll. has published the measurement $\Gamma_{\eta \to \pi^+\pi^-\gamma}/\Gamma_{\eta \to \pi^+\pi^-\pi^0}= 0.175\pm 0.007\pm0.006$, based on 859 $\eta \to \pi^+ \pi^- \gamma$ events \cite{:2007ppa}, which is more then $3\sigma$ below the old measurements. KLOE result \cite{:2011gn}, obtained using $558 ~pb^{-1}$, gives $\Gamma_{\eta \to \pi^+\pi^-\gamma}/\Gamma_{\eta \to \pi^+\pi^-\pi^0}= 0.1838\pm0.0005_{stat}\pm 0.0030_{syst}$, in agreement with the latest CLEO evaluation, providing strong evidence in favour of the box anomaly/direct term.

The knowledge of the $\eta$ meson coupling to virtual photons is important for calculation of anomalous magnetic moment of the muon, because the pseudoscalar exchange is the major contribution to the hadron light-by-light scattering \cite{AmelinoCamelia:2010me}.
In the $\eta\rightarrow e^{+}e^{-}e^{+}e^{-}$ process we have conversion decays, which offer the possibility to precisely measure the virtual photon 4-momentum, via the $e^+$ and $e^-$ 4-momenta and we are directly sensitive to the $\eta$ meson transition form factor because there are no hadrons among the  decay products. The first theoretical evaluation dates from 1967 \cite{1967} and predicts a branching ratio $BR(\eta\rightarrow e^{+}e^{-}e^{+}e^{-})=2.59\times10^{-5}$. Double lepton-antilepton $\eta$ decays have been searched by CMD-2 and WASA, obtaining upper limits at level of the theoretical expectation. KLOE has published the first observation of the $\eta \rightarrow e^{+}e^{-}e^{+}e^{-}$ decay, analysing 
$1.7 ~fb^{-1}$ and identifing $362 \pm 29$ events which results in a branching ratio of $(2.4	\pm 0.2_{stat+bkg}\pm 0.1_{syst})\times10^{-5}$, in agreement with theoretical predictions \cite{:RB}.\\
Vector-meson-dominance assumption provides good description of photon coupling to hadrons, and, implementing systematic corrections to standard VMD, it correctly describes the $\omega \to \pi^0 \mu^+ \mu^-$ experimental results too.
In this framework deviation from standard VMD for the $\phi \to \eta e^+ e^-$ decay spectrum is predicted. The only existing data available come from SND experiment, which has measured the $M_{ee}$ invariant mass distribution with 213 events \cite{Achasov:2000ne}. KLOE has selected $7000$ $\phi \to \eta e^+ e^-$ with $\eta \to \pi^+ \pi^- \pi^0$ using a sample of $739    
~pb^{-1}$. Preliminary fit to the $M_{ee}$ using decay parametrization from \cite{Landsberg:1986fd} and $F(q^2)$ as from \cite{Achasov:1992ku}, indicates the possibility to reach a $5\%$ error on form factor slope.
  
\subsection{Gamma-gamma Physics}
The coupling of photons to scalar and pseudoscalar mesons brings information on meson's quark structure and can be measured directly in $e^+e^-$ colliders via the 
reaction $e^+e^- \to e^+e^- \gamma^* \gamma^* \to e^+e^- X $.
Using the Weizs$\ddot{a}$cker-Williams approximation \cite{Brodsky:1971ud} to understand main qualitative features of the process, when no cuts are applied to the final state leptons, it is possible to evaluate the event yields:
$N_{eeX} = L_{ee}\int \frac{dF}{dW_{\gamma\gamma}}\sigma_{\gamma\gamma \to X}(W_{\gamma\gamma})dW_{\gamma\gamma}$
the $L_{ee}$ is the integrated luminosity, $W_{\gamma\gamma}$ is the mass of the $\gamma^* \gamma^*$ and $dF/dW_{\gamma\gamma}$ the two photons flux function, defined as follows:
$\frac{dF}{dW_{\gamma\gamma}} = \frac{1}{W_{\gamma\gamma}}(\frac{2\alpha}{\pi})^2(ln\frac{E_b}{m_e})^2f(z)$
with $E_{b}$ beam energy and $f(z)$ is a function of $z=\frac{W_{\gamma\gamma}}{2E_b}$. \\
Single $\pi^0$ or $\eta$ production is accessible and this allows to improve determination of two photon decay width of these meson. 
In particular KLOE is looking for $e^+e^-\to e^+e^-\eta$ with $\eta \to \pi^+\pi^-\pi^0$ final state: in a preliminary analysis of 240 $pb^{-1}$ off-peak data about $600$ events from $\eta$ meson, produced in $\gamma \gamma$ interactions  have been disentangled, versus other processes, with a statistical accuracy on $\Gamma_{\gamma\gamma}$ comparable with existing measurements. The same off-peak data have been analysed to search for $e^+e^-\to e^+e^-\eta$ with $\eta \to \pi^0\pi^0\pi^0$ final state.\\
The question concerning $\sigma /f_0(600)$ meson has been debated for a long time. An indirect evidence comes from $\phi \to \pi^0\pi^0 \gamma$ KLOE analysis \cite{Ambrosino:2006hb}.
The $e^+e^- \to e^+e^- \pi^0 \pi^0 $ process is a clean electromagnetic probe to investigate the question, because it is expected to be plainly affected by $\sigma$ contribution.
Our preliminary analysis on the off-peak data,
shows a clear enhancement over estimated backgrounds at low $M_{4\gamma}$; see Fig.\ref{Fig:ggflux}.
\begin{figure}[htb]
  \begin{center}
    \includegraphics[width=0.3\textwidth]{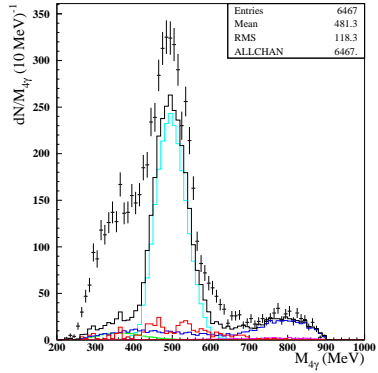}
    \caption{Preliminary spectrum of 4-photons invariant mass with KLOE: dots are data, black line is total MC background,  lightblue is $K_S$ $K_L$ decays, blue is $e^+ e^- \to \omega \pi^0$ and magenta is $\phi \to f_0 \gamma$ ; a clear evidence of $e^+e^- \to e^+ e^- \pi^0 \pi^0$ is given by the excess events at low $M_{4 \gamma}$ invariant mass is visible}
    \label{Fig:ggflux}
  \end{center}
\end{figure}
%
Background subtraction and study of differential cross section together with the understanding of the $\sigma \to\pi\pi$ contribution are in progress.\\
Due to large background from $e^+e^- \to \gamma \gamma (\gamma)$, information from $e^{\pm}$ taggers already installed at KLOE-2, will be crucial in the analysis of new data to look for the production of $\sigma$.\\


The KLOE experiment with $2.5 ~fb^{-1}$ integrated luminosity at the peak of the $\phi$ resonance at the $e^+e^-$ collider DA$\Phi$NE, has published several interesting results. In the next future a new data-taking campaign will be realized by KLOE-2 at the upgraded DA$\Phi$NE, with the aim to collect about $20 ~fb^{-1}$ in order to refine and extend the KLOE physics program. 





%

}  


\end{document}